\documentclass{article}

\usepackage{arxiv}

\usepackage[utf8]{inputenc} 
\usepackage[T1]{fontenc}    
\usepackage{hyperref}       
\usepackage{url}            
\usepackage{booktabs}       
\usepackage{amsfonts}       
\usepackage{nicefrac}       
\usepackage{microtype}      
\usepackage{lipsum}
\usepackage{graphicx}
\usepackage{wrapfig}
\usepackage{multirow}
\usepackage{tcolorbox}
\usepackage{array} 
\tcbuselibrary{breakable}
\graphicspath{ {./images/} {./asset/} }

\title{SIADAFIX: Issue Description Response for Adaptive program repair}

\author{
 XinCao \\
  Code Intelligence Team\\
  LI AUTO\\
  Beijing, China \\
  \texttt{caoxin13@lixiang.com} \\
   \And
 NanYu \\
  Code Intelligence Team\\
  LI AUTO\\
  Beijing, China \\
  \texttt{yunan@lixiang.com} \\ 
}

\begin{document}
\maketitle
\begin{abstract}
We propose utilizing fast and slow thinking to enhance the capabilities of large language model-based agents on complex tasks such as program repair. 
In particular, we design an adaptive program repair method based on issue description response, called SIADAFIX.
The proposed method utilizes slow thinking bug fix agent to complete complex program repair tasks, 
and employs fast thinking workflow decision components to optimize and classify issue descriptions, 
using issue description response results to guide the orchestration of bug fix agent workflows. 
SIADAFIX adaptively selects three repair modes, i.e., easy, middle and hard mode, based on problem complexity.
It employs fast generalization for simple problems and test-time scaling techniques for complex problems.
Experimental results on the SWE-bench Lite show that the proposed method achieves 60.67\% pass@1 performance using the Claude-4 Sonnet model, reaching state-of-the-art levels among all open-source methods. 
SIADAFIX effectively balances repair efficiency and accuracy, providing new insights for automated program repair.
Our code is available at https://github.com/liauto-siada/siada-cli.
\end{abstract}


\section{Introduction}

Recently, Large Language Model (LLM)-based agents ~\cite{wang2024openhands, gao2025trae, yang2024swe, devin2024} with environmental interaction capabilities 
have made progress in the field of code generation. 
The application of agents to complex code tasks such as repository-level program repair ~\cite{jimenez2023swe} has attracted extensive exploration. 
To solve complex programming tasks, some code agents ~\cite{yang2024swe, refact2024} have been endowed with multi-step iteration and deep thinking capabilities. As autonomous systems, these agents use LLMs to iteratively reason about a task, invoke existing tools, then adapt based on outputs produced by these tools~\cite{bouzenia2025understanding}. However, when facing complex tasks like repository-level program repair with inconsistent problem descriptions, existing agents with deep thinking capabilities still cannot stably fix some of the bugs and issues. This may be because allowing models to slowly iterate autonomously leads to continuous error accumulation, making it difficult to break out of inertial thinking.

In "Thinking, Fast and Slow", Kahneman ~\cite{kahneman2011thinking} proposed that human thinking is divided into two modes, i.e., not only slow thinking that requires deliberate iteration, 
but also fast thinking that can make instant decisions based on intuition and accumulated experience. 
Inspired by the above dual-process theory, 
we investigate whether the process steps of program repair can be combined with fast and slow    thinking modes. 
For example, when software experts deal with software bugs, 
they often quickly provide solutions based on experience, 
and then slowly debug step by step according to the solution to finally complete the repair.

Specifically, software experts can quickly refine and expand fault phenomena and their descriptions based on experience when dealing with faults, 
and then gradually complete fault reproduction and localization, patch generation, and patch verification and selection processes based on general fault repair processes and test-driven development principles ~\cite{winter2022developers}. 
The above process embodies the fast and slow thinking modes of human task processing. 
Corresponding to the fast thinking based on experience and intuition, 
we propose using  single LLM requests to refine and expand issue problem descriptions and optimize these descriptions. At the same time, to break through inertial thinking, we propose patch checking to quickly judge repair effectiveness. Corresponding to the slow thinking mode that requires multi-step iteration and interaction with the environment, we designed a bug fix agent with efficient interactive tools to achieve closed-loop thinking for reproduction, localization, patch generation, and testing completion. 
Our design aims to mimic the working mode of human fast and slow thinking to enhance agent capabilities.
\begin{wrapfigure}{r}{0.60\textwidth}
  \centering
\includegraphics[width=0.48\textwidth]{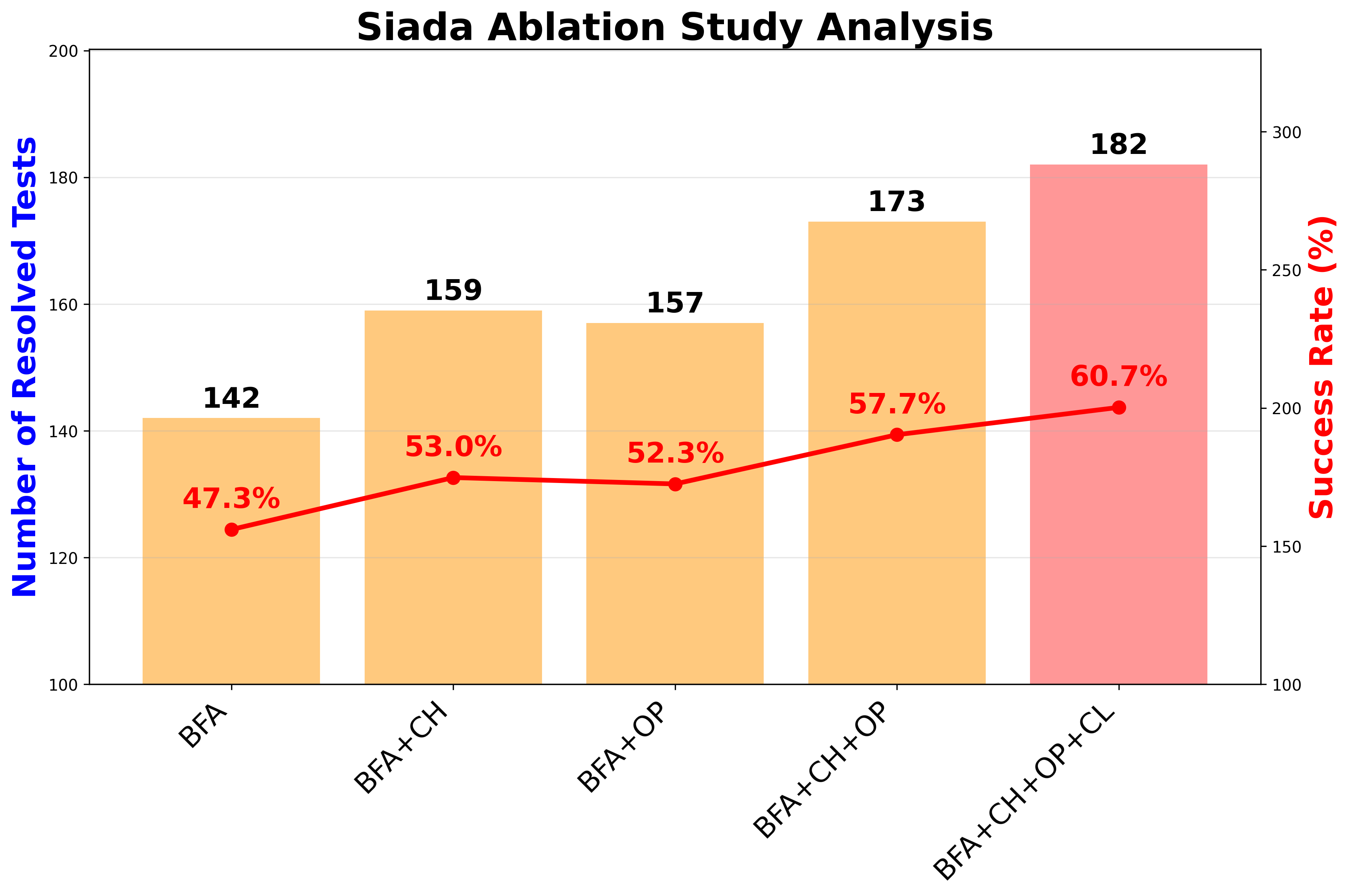}
  \caption{Ablation study of SIADAFIX. BFA: slow thinking Bug fix agent; BFA+CH: BFA enhanced by fast thinking checker; BFA+OP: BFA with fast thinking issue description Optimizer; BFA+CH+OP: BFA combined with issue description Optimizer and checker in single workflow mode; BFA+CH+OP+CL: 
  The proposed SIADAFIX
  including adaptive Classifier (CL) for three modes, i.e. Easy, Middle and Hard.}
  \label{fig:ablation}
\end{wrapfigure}

In fact, we propose SIADAFIX, an adaptive program repair method based on issue description response. 
The core idea of SIADAFIX is to combine bug fix agent with fast and slow thinking modes.
SIADAFIX utilizes slow thinking bug fix agent to complete complex program repair tasks,
and employs fast thinking to optimize and classify issue descriptions,
using issue description optimization and classification response results to guide the orchestration of bug fix agent workflows.
The above approach is beneficial for employing fast generalization strategies for simple problems and multi-step iterative reasoning-time scaling techniques for complex problems. 
Specifically, the proposed method includes two key phases: (i) Workflow decision phase, consisting of an issue description optimizer and issue description classifier; 
(ii) Workflow execution phase, composed of bug repair agents, checkers, and selectors, which can be flexibly combined to form three slow thinking execution workflows, i.e., easy, middle, and hard modes. 
As shown in Figure~\ref{fig:ablation}, we present the ablation experiments results of the proposed fast and slow thinking modules on SWE-Bench Lite @Pass1. 
The ablation study demonstrates the incremental contribution of each component, i.e.,
the basic Bug Fix Agent (BFA) achieves 47.3\% success rate, 
while adding Checker (CH) and Optimizer (OP) components progressively improves performance to 53.0\% and 52.3\% respectively. 
The complete SIADAFIX framework achieved a resolve rate of 182/300 (60.7\%).

Our main contributions include:
\begin{itemize}
\item Based on the human working mode of fast and slow thinking, we propose an adaptive program repair framework based on issue description response;
\item We design issue description optimizer and classifier that achieve adaptive orchestration of easy, middle, and hard repair mode workflows;
\item We utilize various efficient tools and prompt strategies to enhance the capabilities of bug fix agent;
\item Comprehensive evaluation is conducted on the SWE-bench Lite dataset, achieving 60.7\% pass@1 performance using the Claude-4 Sonnet model, reaching state-of-the-art levels among all open-source methods.
\end{itemize}

The remainder of this paper is organized as follows: Section 2 reviews related work; Section 3 details the proposed method; Section 4 presents experimental results and analysis; Section 5 concludes the paper and discusses future work directions.

\section{Related Work}

LLM-based agents~\cite{bouzenia2025understanding, wang2024openhands, gao2025trae, yang2024swe, devin2024, mu2025experepair, xia2024agentless} have gradually become popular in software engineering fields such as code generation and program repair. According to whether they adopt autonomous multi-round interaction with the environment, existing methods can be divided into two major categories: fast thinking with single or hard-coded LLM calls and slow thinking with autonomous planning and LLM invocation.
Fast thinking methods~\cite{jiang2023impact, xia2024agentless, mu2025experepair,xia2024automated} typically operate by prompting a LLM with a query, either a single time or in a hard-coded feedback loop. For example, Agentless achieves program repair localization, repair, and patch verification through hard-coded tool and LLM calls, achieving high performance. Agentless demonstrates the potential of fast thinking mode. Based on this, EXPEREPAIR proposes a memory system that utilizes repair trajectories to optimize prompts in hard-coded processes, further improving fast thinking performance.
Unlike fast thinking methods, which query LLMs with fixed prompt templates and within hard-coded algorithms, slow thinking methods~\cite{wang2024openhands, gao2025trae, yang2024swe, devin2024} typically iteratively reason about a problem, invoke tools, and orchestrate and make decisions about the future based on the output of the previous step. 
For example, SWE-Agent ~\cite{yang2024swe} proposes the Agent-Computer Interface (ACI), utilizing tool set adaptation for language model agents to promote agent-environment interaction and enhance slow thinking capabilities. 
Refact.ai~\cite{refact2024} even constructs deep-analysis tools, forcing agents to engage in more interaction. These methods achieve good results through slow thinking-like modes.
However, the above methods still cannot stably solve some complex problems, possibly because complete reliance on autonomous multi-round iteration by models leads to error accumulation, making it difficult to break out of inertial thinking.
Different from the above methods, 
the proposed SIADAFIX method combines both fast thinking and slow thinking modes, 
utilizing fast thinking to optimize and classify problems, 
thereby guiding slow thinking workflow orchestration, 
improving the efficiency and accuracy of program repair.

\section{Method}
\begin{figure}[h]
  \centering
  \includegraphics[width=1\textwidth]{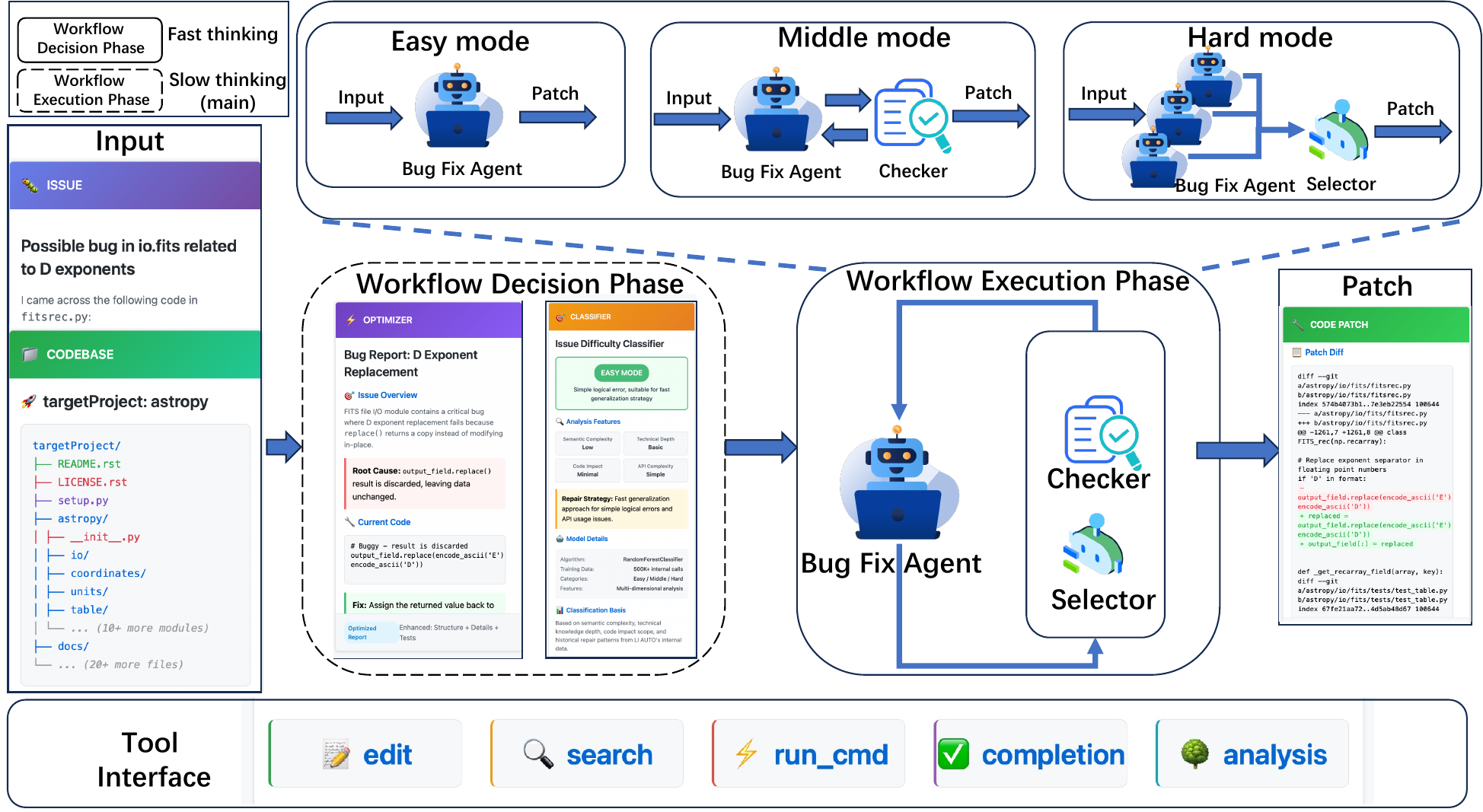}
  \caption{Overview of the SIADAFIX framework. The framework consists of two main phases: (1) Workflow Decision Phase with Issue Description Optimizer and Classifier, and (2) Workflow Execution Phase with bug fix agent, Checkers, and Selectors operating in Easy, Middle, and Hard modes.}
  \label{fig:framework}
\end{figure}

The proposed SIADAFIX method is an adaptive program repair framework based on issue description response. 
By combining fast and slow thinking modes, SIADAFIX improves the efficiency and accuracy of program repair through rapid optimization and classification of problems, guiding slow thinking workflow orchestration. 
As shown in Figure~\ref{fig:framework}, Based on tools of SIADA-CLI, input issue descriptions (problem statement), and target codebase, SIADAFIX completes issue description optimization and repair mode selection through the workflow decision phase, then utilizes the designed bug fix agent, checker, and selector in the workflow execution phase to complete repair and output corresponding code patches.

\subsection{Workflow Decision Phase}
When an issue description is input, SIADAFIX first enters the workflow decision phase, which is responsible for analyzing the input issue description and deciding which repair strategy to adopt based on the analysis results. This phase contains two core components: issue description optimizer and the issue description difficulty classifier.

\subsubsection{Issue Description Optimizer}

In software development, bug issue descriptions provide crucial information for developers. Compared to SWE-bench~\cite{jimenez2023swe} Lite, existing methods achieve higher scores more easily on the SWE-bench Verified leaderboard. For example, Refact.ai Agent\cite{refact2024} achieved 74.4\% pass@1 problem resolution rate on Verified, but only 60\% on Lite. This may be because problem descriptions in SWE-bench Verified are manually annotated, such as marking and distinguishing the clarity and specificity of problem description tasks This inspires us to think about how to make models automatically "annotate" problem descriptions. 
Meanwhile, pre-trained LLMs contain extensive software engineering knowledge~\cite{hou2024large}. 
Leveraging this knowledge to improve inconsistent issue descriptions may enhance intelligent agents' bug repair capabilities in real-world scenarios. 

First, through experimental analysis and research, 
we found that reproduction steps, stack traces, test cases, and acceptance criteria in bug descriptions are useful content. 
Based on this, we propose that the goal of the issue description optimizer is to extract and supplement key information from original descriptions to generate structured issue reports. 
Specifically, the optimizer optimizes problem descriptions in four aspects: problem reproduction step extraction, expected behavior clarification, error phenomenon summarization, and acceptance criteria definition. 
The problem reproduction step extraction process identifies and extracts specific operational steps that lead to bugs from descriptions, including input data, execution environment, call sequences, etc. The expected behavior clarification process analyzes and clarifies the program's expected behavior, including correct output results, logical branches that should be executed, etc. Error phenomenon summarization summarizes current error phenomena, including exception information, error output, program crashes, etc. Acceptance criteria definition defines acceptance criteria for successful repair based on problem descriptions, providing a basis for subsequent repair verification.

The optimizer adopts large language model-based information extraction technology, guiding models to extract structured information from unstructured issue descriptions through carefully designed prompt templates. The optimized issue reports not only contain richer technical details but also provide standardized input formats for subsequent classification and repair. Specific details are provided in Appendix~\ref{appendix:optimizer}.

\subsubsection{Issue Description difficulty Classifier}

In software development and maintenance processes, development teams receive numerous issue tickets through defect tracking systems (such as Jira\cite{valdez2020sentiment}, Bugzilla\cite{serrano2005bugzilla}). 
These tickets contain various real bugs, feature improvement requirements, code refactoring suggestions, performance optimization issues, etc. 
Researchers have proposed automatic classification\cite{valdez2020sentiment, pandey2017automated} through issue descriptions (issue tickets) to achieve automated task allocation and development cycle planning, 
which is crucial for efficient bug handling. 
Inspired by this, we propose a bug issue description classifier, 
which is a RandomForestClassifier\cite{pedregosa2011scikit} responsible for evaluating problem complexity. Based on over 500,000 calls to LI AUTO's internal code generation tools from April to September 2025, we extracted data similar to issue descriptions and conducted analysis, classifying issue descriptions into Easy, Middle, and Hard levels.

The classifier adopts machine learning models based on text analysis, combined with rule engines to achieve accurate complexity assessment. 
The classification results directly determine the subsequent repair strategies, including easy, middle, hard mode. 
Easy model suites for simple logical errors, API usage errors, etc., adopting fast generalization strategies.
Middle mode suites for medium complexity problems, adopting iterative optimization strategies. 
Hard mode suites for complex systemic problems, adopting multi-candidate generation and selection strategies.

The training data for this classifier is extracted from our internal code generation requirement data, which is similar to problem descriptions in the Lite dataset~\cite{jimenez2023swe}. 
The extracted relevant features are shown in the following table, and we have open-sourced this model weights in SIADA CLI~\cite{siadacli2024}. 
The classification decision is based on multi-dimensional feature analysis, e.g., Semantic Complexity Analysis analyzes semantic features such as concept complexity and logical relationship complexity involved in issue descriptions; Technical Difficulty Assessment evaluates the depth of technical knowledge required for repair, including the number of APIs involved, framework complexity, algorithm difficulty, etc. Code Impact Scope estimates the code scope that repair may affect, including the number of files, functions, inter-module dependencies, etc. Historical Repair Data utilizes patterns in historical repair data to assist classification decisions, as well as sentiment tone words related to issue difficulty. The specific feature calculation methods are shown in Appendix~\ref{appendix:features}.

\subsection{Workflow Execution Phase}

Benefiting from the optimization and classification results of the workflow decision phase, 
SIADAFIX possesses relatively clear problem descriptions and execution strategies in the workflow execution phase. 
This phase adopts corresponding repair strategies to execute specific program repair tasks based on the output of the decision phase. This phase accomplishes different repair strategies for easy, middle, and hard modes through flexible combination of the designed bug fix agent, checker, and selector.

\subsubsection{Bug Fix Agent}
The bug fix agent is the core component for executing specific repair tasks, implemented based on large language models. 
Meanwhile, Based on previous open-source work\cite{cline2024, aider2024, yang2024swe, wang2024openhands}, we have implemented efficient tools to help the bug fix agent better interact with the codebase and other environments, including run cmd tools that support Mac, Windows, and Linux operating systems simultaneously, as well as high-performance search tools, editing tools that support querying and modifying files, and AST tools that can analyze code syntax structures.
We have conducted extensive optimization of these tools for code generation tasks. For example, in the search tool, the number of files to search is set to 300, and when search file segments are output without line numbers, it can reduce interference to the LLM.

\begin{table}[h]
\centering
\caption{Tools integrated in the Bug Fix Agent for enhanced program repair capabilities.}
\label{tab:tools}
\resizebox{\textwidth}{!}{%
\begin{tabular}{>{\centering\arraybackslash}m{3cm}>{\centering\arraybackslash}m{4cm}>{\centering\arraybackslash}m{8cm}}
\toprule
\textbf{Tool} & \textbf{Purpose} & \textbf{Key Features} \\
\midrule
edit & File Content Modification & Multi-format support with base64 encoding; Multiple edit operations (view, create, str\_replace, insert); Error handling with UTF-8 validation and permission checks; Context-aware editing with line-based operations \\
\midrule
regex\_search\_files & High-Performance Code Search & Ripgrep integration for ultra-fast searching; Cross-platform binary management with automatic detection; Advanced regex support with Unicode; Context-aware results with before/after context lines; Smart file filtering with glob pattern support \\
\midrule 
run\_cmd & System Command Execution & Environment adaptation between pexpect (Unix) and subprocess (Windows); Real-time output streaming with error handling; Working directory management; Exit code tracking with detailed reporting \\
\midrule
fix\_attempt\_completion & Task Completion Validation & Mandatory completion enforcement; Detailed reporting with comprehensive fix summary; Status tracking with clear completion indicators; Workflow validation ensuring all bugs are addressed \\
\midrule
list\_code\_\newline definition\_names & Advanced AST Code Analysis & Tree-sitter integration with 40+ language support; Dual extraction modes for definitions and references; Smart query system with language-specific files; Contextual tree generation with hierarchical structure \\
\bottomrule
\end{tabular}%
}
\end{table}

\begin{figure}[h]
  \centering
  \includegraphics[width=1\textwidth]{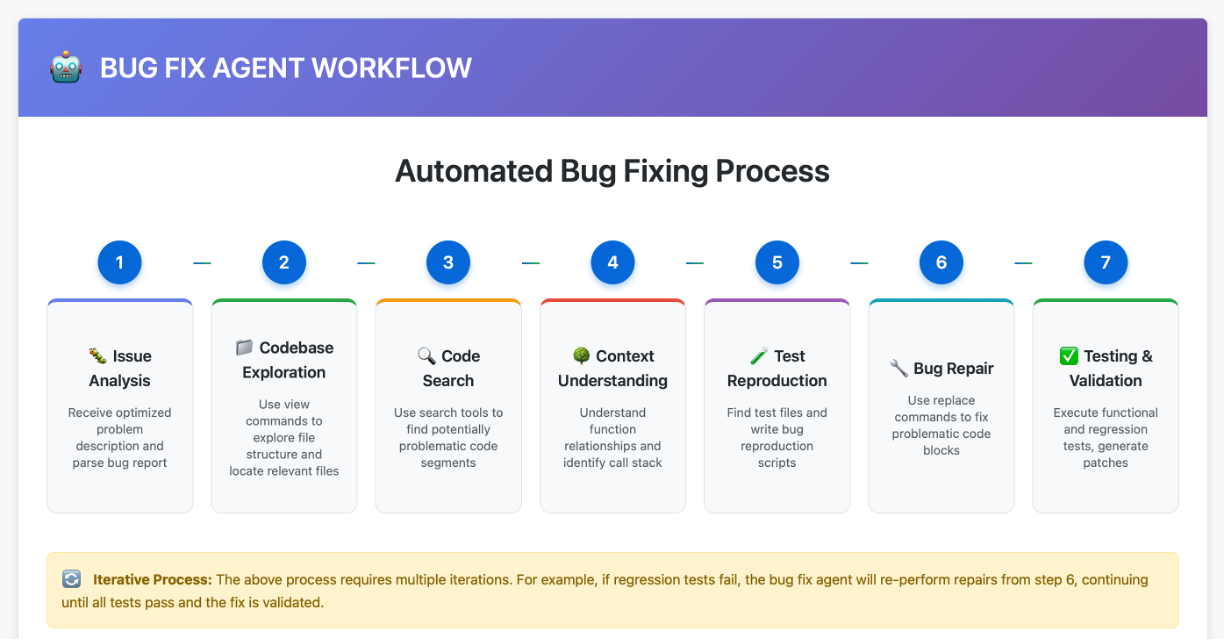}
  \caption{Overview of the Bug FIX Agent.}
  \label{fig:BugFixAgent}
\end{figure}

Specifically, the proposed bug fix agent is designed as a programming expert (detailed in the appendix), equipped with various tools to analyze, locate, and repair user-submitted tasks or problems.
The specific steps for solving a problem are show in Figure~\ref{fig:BugFixAgent}. After obtaining the optimized problem description from the optimizer, the bug fix agent first explores the codebase structure, for example, using the view command in editing tools to examine the secondary file and folder structure under the project root directory, iterating until locating program files related to the problem. For problem-related program files, the bug fix agent uses search tools to find potentially problematic code segments. Interestingly, through experiments, we found that search tool code segment outputs without line numbers can reduce LLM interference, while the view command in editing tools requires line numbers to help LLMs read files more stably. We believe this may be because the line numbers in multiple code segment outputs from search are relatively random, while the view command generally reads sequentially from the first line of the file, and these continuous line numbers have strong sequential relationships that benefit LLM judgment. After locating the problematic code segment, the bug fix agent reads nearby or semantically similar contextual code, understands function call relationships, and identifies the call stack that the current problem may involve. Meanwhile, it uses search tools to find existing test files related to relevant functions or classes in the problem project, and based on these files, completes the writing of bug reproduction scripts. After completing problem reproduction, based on the reproduction output, it uses the replace command in editing tools to repair the buggy code blocks, then performs various functional and regression tests to ensure the correctness of the repair. notability, The above process requires multiple iterations, for example, if regression tests fail, the bug fix agent will re-perform repairs. Finally, after completing problem repair with completion tool, it provides a repair summary and generates the corresponding patch diff through git tools.

\subsubsection{Multi-level Checker System}

Based on previous multi-agent experiments, we found that if subsequent agents continue the work of previous agents, they must trust the work of previous agents to some extent, which creates inertial thinking. If there are problems with previous agents, all subsequent work becomes meaningless. 
Therefore, we believe there needs to be someone standing on the opposite side of the bug repair agent, forgetting the bug repair agent's work process, and then reviewing its repair results. Thus, we added a patch checking phase.

To ensure repair quality, we designed a multi-level checker system, including a checker and an enhanced checker. A checker that takes issue descriptions and code patches provided by Bug Fix Agent as input, deciding whether to proceed to the next round of repair. 
An enhanced checker that, compared to the basic checker, adds Bug Fix Agent's problem-solving trajectory as input, providing deeper quality assessment including code style, performance impact, potential side effects, etc.

We verify whether patch repairs are consistent with problem descriptions to determine whether bug repair needs to be re-executed—this avoids unnecessary repeated model calls and saves computational resources. The checking process is divided into two stages: (i) deciding whether to re-execute bug repair based solely on problem descriptions; (ii)  based on problem descriptions and execution trajectories from previous bug repair iterations, providing more targeted suggestions to help the next bug repair cycle solve problems more precisely.

Actually, the reason for adding this checking stage is that the model's original workflow lacks independent self-verification capabilities. It often ends directly after completing one repair output, lacking comprehensive comparison between repair results and requirement descriptions, easily missing details or introducing regression errors. Additional checking can independently verify parts not adequately covered by the bug fix stage's reasoning path, avoiding complete dependence on previous repair trajectories that may produce path dependencies, thereby improving the detection rate of missed problems and repair quality.

\subsubsection{Selector}
The Selector is designed to select the best repair result from multiple patch diffs generated by the bug fix agent.
It tests and compares multiple patch diffs, selecting the one that best meets the problem description and passes all tests as the final output. 
The selector uses a voting mechanism to evaluate each patch diff based on criteria such as relevance, completeness, safety, and code quality, ultimately selecting the optimal patch diff. Specifically, relevance represents whether the patch directly addresses the original issue; completeness evaluates whether the patch provides a complete solution; safety assesses whether the patch avoids introducing new problems; and code quality measures the maintainability and readability of the patch code.

\subsubsection{Three Repair Modes}
The agent adopts different repair strategies according to different complexity levels. Through analysis of real-world code generation requirements and descriptions, we believe that some simple problems in problem descriptions can be solved through relatively concise workflows, corresponding to the easy mode of SIADAFIX. Therefore, when problems are classified as simple problems, SIADAFIX uses a two-stage workflow of issue description optimization → issue repair to solve them. It adopts a single-round repair approach, directly generating repair patches based on optimized issue descriptions.

For another part of problems, we believe that models may overlook important boundary conditions during repair, corresponding to the middle mode of SIADAFIX, which uses a three-stage workflow of issue description optimization → issue repair → patch checking. This is a multi-round iterative repair approach, where each round includes a complete cycle of repair generation, verification, and feedback.

Finally, for problems with relatively vague descriptions or inherently difficult problems, corresponding to hard mode. After issue description optimization, we utilize multiple issue repairs to obtain multiple results, and use a selector to choose a more suitable result through testing and comparison. Hard mode adopts a multi-candidate generation strategy, generating multiple repair candidates in each round, and resetting the working environment after each iteration to avoid cumulative effects of erroneous repairs. In hard mode, we designed an agent for selecting multiple code patches, which selects the optimal result by testing and comparing multiple patch diff results from the bug fix agent.

\section{Experiments}

\subsection{Experimental Setup}

\subsubsection{Dataset}

We evaluates the performance of the proposed method on the SWE-bench Lite\cite{jimenez2023swe} dataset. SWE-bench Lite is a benchmark dataset specifically designed for evaluating automated program repair methods, containing 300 real software problems from popular GitHub repositories. These problems cover different programming languages, frameworks, and problem types, providing good representativeness.

The SWE-bench Lite dataset has good authenticity and representativeness, with all problems coming from real software defects in popular open-source projects such as Django, Flask, NumPy, Pandas, etc. Each problem contains complete problem descriptions, reproduction steps, and expected repair results, and provides an automated testing framework for verifying repair effectiveness, ensuring the objectivity and reproducibility of experimental evaluation. This paper adopts Pass@1 as the main evaluation metric.

\subsubsection{Baselines}

For repository-level program repair, academia and industry have conducted extensive research. Many methods have been evaluated on SWE-bench, with Claude-4 Sonnet-based methods showing the best performance. We mainly focus on and compare open-source methods among them. The experiments compared 4 methods using Claude-4 Sonnet model on the SWE-bench leaderboard, 
including ExpeRepair-v1.0~~\cite{mu2025experepair}, Refact.ai Agent~~\cite{refact2024}, KGCompass~~\cite{yang2025enhancing}, SWE-agent~~\cite{yang2024swe}, etc.
     
\subsection{Main Results}

\subsubsection{Overall Performance}

Table~\ref{tab:main_results} presents the comprehensive comparison of our proposed SIADAFIX method against state-of-the-art baselines on the SWE-bench Lite dataset. The results demonstrate the effectiveness of our adaptive approach across different project types and complexity levels.

\begin{table}[h]
\centering
\caption{Main experimental results on SWE-bench Lite dataset. Numbers show resolved/total problems for each project.}
\label{tab:main_results}
\resizebox{\textwidth}{!}{%
\begin{tabular}{l|c|c|c|c|c|c}
\toprule
\multirow{2}{*}{\textbf{Project}} & \multirow{2}{*}{\textbf{SIADAFIX}} & \multirow{2}{*}{\textbf{SWE-agent}} & \multirow{2}{*}{\textbf{KGCompass}} & \multirow{2}{*}{\textbf{ExpeRepair-v1}} & \textbf{Refact.ai} & \multirow{2}{*}{\textbf{Total Problems}} \\
 & & & & & \textbf{Agent} & \textbf{} \\
\midrule
django/django & 77/114 (67.5\%) & 75/114 (65.8\%) & \textbf{79/114 (69.3\%)} & \textbf{79/114 (69.3\%)} & 78/114 (68.4\%) & 114 \\
sympy/sympy & \textbf{44/77 (57.1\%)} & 40/77 (51.9\%) & 40/77 (51.9\%) & 41/77 (53.2\%) & 43/77 (55.8\%) & 77 \\
scikit-learn/scikit-learn & \textbf{17/23 (73.9\%)} & 15/23 (65.2\%) & 15/23 (65.2\%) & 16/23 (69.6\%) & \textbf{17/23 (73.9\%)} & 23 \\
matplotlib/matplotlib & \textbf{14/23 (60.9\%)} & 9/23 (39.1\%) & 11/23 (47.8\%) & 12/23 (52.2\%) & 11/23 (47.8\%) & 23 \\
pytest-dev/pytest & 10/17 (58.8\%) & 10/17 (58.8\%) & \textbf{11/17 (64.7\%)} & 10/17 (58.8\%) & 10/17 (58.8\%) & 17 \\
sphinx-doc/sphinx & \textbf{9/16 (56.3\%)} & 8/16 (50.0\%) & 2/16 (12.5\%) & \textbf{9/16 (56.3\%)} & 6/16 (37.5\%) & 16 \\
pylint-dev/pylint & 3/6 (50.0\%) & 3/6 (50.0\%) & 3/6 (50.0\%) & \textbf{4/6 (66.7\%)} & 3/6 (50.0\%) & 6 \\
psf/requests & 2/6 (33.3\%) & 3/6 (50.0\%) & \textbf{5/6 (83.3\%)} & 3/6 (50.0\%) & \textbf{5/6 (83.3\%)} & 6 \\
astropy/astropy & 3/6 (50.0\%) & 3/6 (50.0\%) & 3/6 (50.0\%) & 3/6 (50.0\%) & 3/6 (50.0\%) & 6 \\
pydata/xarray & 1/5 (20.0\%) & 2/5 (40.0\%) & \textbf{3/5 (60.0\%)} & 2/5 (40.0\%) & 2/5 (40.0\%) & 5 \\
mwaskom/seaborn & 2/4 (50.0\%) & 2/4 (50.0\%) & 2/4 (50.0\%) & 2/4 (50.0\%) & 2/4 (50.0\%) & 4 \\
pallets/flask & 0/3 (0.0\%) & 0/3 (0.0\%) & \textbf{1/3 (33.3\%)} & 0/3 (0.0\%) & 0/3 (0.0\%) & 3 \\
\midrule
{Total} & \textbf{182/300 (60.7\%)} & {170/300 (56.7\%)} & {175/300 (58.3\%)} & {181/300 (60.3\%)} & {180/300 (60.0\%)} & {300} \\
\bottomrule
\end{tabular}%
}
\end{table}

The experimental results show that our proposed SIADAFIX method achieves 60.7\% Pass@1 performance on the SWE-bench Lite dataset, successfully solving 182 out of 300 problems. This represents a competitive performance among existing methods, with SIADAFIX outperforming SWE-agent by 4.0 percentage points (182 vs. 170 problems solved) and KGCompass by 2.4 percentage points (182 vs. 175 problems solved). SIADAFIX demonstrates particularly strong performance on several projects: achieving the best results on sympy/sympy (44 problems solved, 57.1\%), matplotlib/matplotlib (14 problems solved, 60.9\%), and sphinx-doc/sphinx (9 problems solved, 56.3\%). The method shows consistent performance across different project types, from web frameworks like Django to scientific computing libraries like scikit-learn, showcasing the effectiveness of our adaptive workflow orchestration approach.

\subsubsection{Ablation Study}

To validate the effectiveness of each component in our proposed SIADAFIX framework, we conduct comprehensive ablation experiments. Table~\ref{tab:ablation} shows the performance of different component combinations across all projects on the SWE-bench Lite dataset.

\begin{table}[h]
\centering
\caption{Ablation study results on SWE-bench Lite dataset by project. BFA: Bug Fix Agent; CH: Checker; OP: Optimizer; CL: Classifier.}
\label{tab:ablation}
\resizebox{\textwidth}{!}{%
\begin{tabular}{l|c|c|c|c|c|c}
\toprule
\multirow{2}{*}{\textbf{Project}} & \multirow{2}{*}{\textbf{BFA}} & \multirow{2}{*}{\textbf{BFA+CH}} & \multirow{2}{*}{\textbf{BFA+OP}} & \multirow{2}{*}{\textbf{BFA+CH+OP}} & \textbf{BFA+CH+OP+CL} & \multirow{2}{*}{\textbf{Total Problems}} \\
 & & & & & \textbf{(SIADAFIX)} & \textbf{} \\
\midrule
astropy/astropy & 1 (16.7\%) & 3 (50.0\%) & 3 (50.0\%) & 3 (50.0\%) & 3 (50.0\%) & 6 \\
django/django & 60 (52.6\%) & 65 (57.0\%) & 67 (58.8\%) & 72 (63.2\%) & 77 (67.5\%) & 114 \\
matplotlib/matplotlib & 8 (34.8\%) & 10 (43.5\%) & 11 (47.8\%) & 13 (56.5\%) & 14 (60.9\%) & 23 \\
mwaskom/seaborn & 2 (50.0\%) & 2 (50.0\%) & 1 (25.0\%) & 2 (50.0\%) & 2 (50.0\%) & 4 \\
pallets/flask & 0 (0.0\%) & 0 (0.0\%) & 0 (0.0\%) & 0 (0.0\%) & 0 (0.0\%) & 3 \\
psf/requests & 1 (16.7\%) & 2 (33.3\%) & 1 (16.7\%) & 2 (33.3\%) & 2 (33.3\%) & 6 \\
pydata/xarray & 0 (0.0\%) & 0 (0.0\%) & 0 (0.0\%) & 0 (0.0\%) & 1 (20.0\%) & 5 \\
pylint-dev/pylint & 2 (33.3\%) & 2 (33.3\%) & 3 (50.0\%) & 3 (50.0\%) & 3 (50.0\%) & 6 \\
pytest-dev/pytest & 7 (41.2\%) & 8 (47.1\%) & 8 (47.1\%) & 9 (52.9\%) & 10 (58.8\%) & 17 \\
scikit-learn/scikit-learn & 13 (56.5\%) & 15 (65.2\%) & 14 (60.9\%) & 16 (69.6\%) & 17 (73.9\%) & 23 \\
sphinx-doc/sphinx & 6 (37.5\%) & 7 (43.8\%) & 7 (43.8\%) & 8 (50.0\%) & 9 (56.3\%) & 16 \\
sympy/sympy & 42 (54.5\%) & 45 (58.4\%) & 42 (54.5\%) & 44 (57.1\%) & 44 (57.1\%) & 77 \\
\midrule
\textbf{Total} & \textbf{142 (47.3\%)} & \textbf{159 (53.0\%)} & \textbf{157 (52.3\%)} & \textbf{173 (57.7\%)} & \textbf{182 (60.7\%)} & \textbf{300} \\
\bottomrule
\end{tabular}%
}
\end{table}

The ablation study demonstrates the incremental contribution of each component across different projects. The basic Bug Fix Agent (BFA) achieves 47.3\% overall success rate, with notable variation across projects - performing best on sympy/sympy (54.5\%) and scikit-learn/scikit-learn (56.5\%). Adding the Checker (CH) consistently improves performance across most projects, achieving 53.0\% overall success rate with significant improvements in django/django (+4.4\%) and scikit-learn/scikit-learn (+8.7\%). The Optimizer (OP) shows similar contribution patterns with 52.3\% overall success rate. The combination BFA+CH+OP achieves 57.7\% success rate, demonstrating synergistic effects particularly evident in matplotlib/matplotlib (+21.7\% over BFA) and django/django (+10.6\% over BFA). Finally, the complete SIADAFIX framework (BFA+CH+OP+CL) with the adaptive Classifier achieves the best performance at 60.7\%, showing consistent improvements across all project types and representing a 13.4 percentage point improvement over the baseline BFA approach.

\section{Conclusion and Future Work}

This paper proposes an adaptive program repair method based on issue description response. The core idea of this method is to achieve adaptive workflow orchestration for the proposed bug fix agent,
checkers, and selectors through intelligent optimization and classification of bug issue descriptions: employing fast
generalization strategies for simple problems and test-time scaling techniques for complex problems.

Based on current research achievements and limitations, future work will focus on three core directions. First, introducing specialized architecture analysis tools by integrating software architecture analysis techniques to enhance understanding of complex systemic problems, which will help handle complex bug repair scenarios involving multiple modules and component interactions. Second, implementing multi-modal information fusion by combining code, documentation, test cases, and other information sources to improve the comprehensiveness of problem understanding, enabling more accurate grasp of problem essence and repair requirements. Finally, conducting domain knowledge integration by building domain-specific knowledge bases to enhance repair capabilities in specific domains, particularly in professional fields such as machine learning, graphics, and databases.

\subsection{Conclusion}

The adaptive program repair method based on issue description response proposed in this paper provides new ideas and solutions for the automated program repair field. Through intelligent problem analysis and adaptive workflow orchestration, this method significantly improves repair efficiency while ensuring repair quality. Experimental results show that the proposed method achieves state-of-the-art performance on the SWE-bench Lite dataset, with important theoretical value and practical application prospects.

With the continuous development of large language model technology and the deepening of software engineering practices, automated program repair technology will play an increasingly important role in software development and maintenance. We believe that through continuous technological innovation and practical exploration, automated program repair will ultimately achieve the transformation from auxiliary tools to core capabilities, bringing revolutionary changes to the software engineering field.

\bibliographystyle{unsrt}
\bibliography{references}  







\appendix
\section{Feature Extraction Methods for Issue Description Classification}
\label{appendix:features}

This appendix provides detailed information about the feature extraction methods used in the issue description difficulty classifier described in the main paper.

\begin{table}[h]
\centering
\caption{Feature extraction methods for issue description classification}
\label{tab:features}
\resizebox{\textwidth}{!}{%
\begin{tabular}{|c|l|l|l|c|c|l|}
\hline
\textbf{No.} & \textbf{Feature Name} & \textbf{Calculation Method} & \textbf{Feature Type} & \textbf{Data Type} & \textbf{Example} & \textbf{Description} \\
\hline
0 & char\_count & Character-level text length measurement & Basic Statistics & int & 1250 & Total character count \\
\hline
1 & word\_count & Token-based text segmentation and counting & Basic Statistics & int & 180 & Total word count \\
\hline
2 & line\_count & Line-break delimiter analysis for text structure & Basic Statistics & int & 15 & Total line count \\
\hline
3 & sentence\_count & Sentence boundary detection using punctuation markers & Basic Statistics & int & 8 & Total sentence count \\
\hline
4 & avg\_word\_length & Statistical mean calculation of lexical unit lengths & Language Complexity & float & 5.2 & Average word length \\
\hline
5 & unique\_word\_ratio & Lexical diversity measurement through vocabulary analysis & Language Complexity & float & 0.75 & Unique word ratio \\
\hline
6 & project\_mentions & Domain-specific terminology identification and quantification & Project Specific & int & 3 & Project-related keywords \\
\hline
7 & error\_mentions & Error-related vocabulary detection and enumeration & Problem Analysis & int & 2 & Error-related keywords \\
\hline
8 & tech\_mentions & Technical terminology frequency analysis & Technical Content & int & 5 & Technical term occurrences \\
\hline
9 & code\_blocks & Pattern matching for inline and block code segments & Code Content & int & 2 & Number of code blocks \\
\hline
10 & code\_pattern\_count & Syntactic code structure identification and enumeration & Code Content & int & 4 & Code pattern occurrences \\
\hline
11 & urls & Pattern matching for HTTP/HTTPS URLs and web addresses & External References & int & 1 & Number of URL links \\
\hline
12 & version\_mentions & Pattern matching for semantic version identifiers & Version Information & int & 2 & Version number occurrences \\
\hline
13 & number\_count & Pattern matching for standalone numeric values & Numeric Content & int & 8 & Independent numbers \\
\hline
14 & sentiment\_score & Affective polarity assessment through lexical analysis & Sentiment Analysis & int & -1 & Sentiment tendency score \\
\hline
15 & question\_count & Interrogative linguistic structure identification & Problem Analysis & int & 3 & Question word occurrences \\
\hline
16 & uppercase\_ratio & Capitalization frequency analysis & Text Style & float & 0.05 & Uppercase letter ratio \\
\hline
17 & punctuation\_ratio & Orthographic symbol density measurement & Text Style & float & 0.08 & Punctuation mark ratio \\
\hline
18 & chars\_per\_word & Character-to-token ratio computation & Derived Metrics & float & 6.9 & Average chars per word \\
\hline
19 & sentences\_per\_line & Sentence density per textual unit analysis & Derived Metrics & float & 0.53 & Average sentences per line \\
\hline
\end{tabular}%
}
\end{table}

\subsection{Feature Categories}

The features are organized into several categories:

\begin{itemize}
\item \textbf{Basic Statistics} (Features 0-3): Fundamental text metrics including character count, word count, line count, and sentence count.
\item \textbf{Language Complexity} (Features 4-5): Metrics that capture the linguistic complexity of the issue description.
\item \textbf{Project Specific} (Feature 6): Keywords related to specific projects or frameworks.
\item \textbf{Problem Analysis} (Features 7, 14-15): Features that help identify the nature and sentiment of the problem.
\item \textbf{Technical Content} (Feature 8): Count of technical terms and jargon.
\item \textbf{Code Content} (Features 9-10): Metrics related to code blocks and patterns in the description.
\item \textbf{External References} (Feature 11): Count of URLs and external links.
\item \textbf{Version Information} (Feature 12): Detection of version numbers and identifiers.
\item \textbf{Numeric Content} (Feature 13): Count of standalone numbers in the text.
\item \textbf{Text Style} (Features 16-17): Stylistic features like capitalization and punctuation usage.
\item \textbf{Derived Metrics} (Features 18-19): Computed ratios and averages from basic statistics.
\end{itemize}

These features are used by the RandomForestClassifier to automatically categorize issue descriptions into Easy, Middle, and Hard difficulty levels, enabling adaptive workflow orchestration in the SIADAFIX framework.

\section{Issue Description Optimizer}
\label{appendix:optimizer}
\begin{tcolorbox}[colback=blue!5!white,colframe=blue!75!black,title=Issue Description Optimizer Prompt]
\small
Please analyze the following bug description and generate a more \textbf{complete} and \textbf{precise} bug report so that developers can fully understand and fix the issue.

\subsection{Optimization Requirements}

\begin{enumerate}
\item \textbf{Identify All Potential Issues}
   \begin{itemize}
   \item Do not only focus on the explicit error mentioned by the user, but also identify root causes that may lead to the error
   \item Analyze technical details exposed in the error message
   \end{itemize}

\item \textbf{Clarify Test Scenarios}

\item \textbf{Define Expected Behavior}
   \begin{itemize}
   \item Clearly describe how different input formats should be handled
   \item Require not only error recovery but also correct handling of all reasonable inputs
   \end{itemize}

\item \textbf{Provide Complete Reproduction Steps}
   \begin{itemize}
   \item Include specific, runnable code examples
   \item Cover multiple data format scenarios
   \end{itemize}

\item \textbf{Define Success Criteria}
   \begin{itemize}
   \item List all conditions that must be satisfied after the fix
   \item Ensure both error recovery and data compatibility are included
   \end{itemize}
\end{enumerate}

\subsection{Principles}

\begin{enumerate}
\item Do not omit or alter any information from the original bug description.
\end{enumerate}

\subsection{Output Format}

Generate a \textbf{structured bug report} that includes:

\begin{itemize}
\item \textbf{Issue Overview}
\item \textbf{Detailed Problem Description} (including root cause)
\item \textbf{Reproduction Steps} (with multiple scenarios)
\item \textbf{Expected Behavior}
\item \textbf{Acceptance Criteria}
\end{itemize}

\subsection{Here is the original bug description:}

Possible bug in io.fits related to D exponents
I came across the following code in \texttt{fitsrec.py}:

\begin{verbatim}
        # Replace exponent separator in floating point numbers
        if 'D' in format:
            output_field.replace(encode_ascii('E'), encode_ascii('D'))
\end{verbatim}

I think this may be incorrect because as far as I can tell \texttt{replace} is not an in-place operation for \texttt{chararray} (it returns a copy). Commenting out this code doesn't cause any tests to fail so I think this code isn't being tested anyway.
\end{tcolorbox}

\section{Bug Fix Agent}

\begin{tcolorbox}[colback=blue!5!white,colframe=blue!75!black,title=Bug Fix Agent Prompt,breakable]
\small
You are Siada, a bug fix agent with extensive knowledge in many programming languages, frameworks, design patterns, and best practices.

\subsection{TOOL USE}

You have access to a set of tools. You can use one tool per message, and will receive the execution results of the tool. You use tools step-by-step to accomplish a given task, with each tool use informed by the result of the previous tool use.

\subsection{CAPABILITIES}

\begin{itemize}
\item You have access to tools that let you execute CLI commands on the user's computer, list files, view source code definitions, regex search, read and edit files. These tools help you effectively accomplish a wide range of tasks, such as writing code, making edits or improvements to existing files, understanding the current state of a project, performing system operations, and much more.
\item You can use search\_files to perform regex searches across files in a specified directory, outputting context-rich results that include surrounding lines. This is particularly useful for understanding code patterns, finding specific implementations, or identifying areas that need refactoring.
\item You can use the list\_code\_definition\_names tool to get an overview of source code definitions for all files at the top level of a specified directory. This can be particularly useful when you need to understand the broader context and relationships between certain parts of the code. You may need to call this tool multiple times to understand various parts of the codebase related to the task.
\item You can use the run\_cmd tool to run commands on the user's computer whenever you feel it can help accomplish the user's task. When you need to execute a CLI command, you must provide a clear explanation of what the command does. Prefer to execute complex CLI commands over creating executable scripts, since they are more flexible and easier to run.
\end{itemize}

\subsection{RULES}

\begin{itemize}
\item Before starting the actual work, please first understand the user's task and make a plan.
\item Your current working directory is: /testbed
\item You cannot cd into a different directory to complete a task. You are stuck operating from '/testbed', so be sure to pass in the correct 'path' parameter when using tools that require a path.
\item Do not use the \textasciitilde{} character or \$HOME to refer to the home directory.
\item You should frequently use the compress\_context\_tool to summarize historical messages, aiming to keep your message history as concise and accurate as possible.
\item When using the regex\_search\_files tool, craft your regex patterns carefully to balance specificity and flexibility.
\item When making changes to code, always consider the context in which the code is being used. Ensure that your changes are compatible with the existing codebase and that they follow the project's coding standards and best practices.
\item Please fix the bug while simultaneously performing comprehensive edge testing to identify and address all boundary conditions, extreme scenarios, exceptional cases, null/empty inputs, maximum/minimum values, invalid data types, concurrent access issues, and resource constraints.
\item After completing the fix, validate that the entire system works correctly under all conditions by running thorough tests on both the original bug and all identified edge cases.
\item When ANY bug fixing task is complete, you MUST call the fix\_attempt\_completion tool. This applies to ALL tasks, even simple ones.
\item You are not allowed to ask questions to the user, generate commands requiring user input, or any other similar interactions. Each task must be completed independently.
\item Avoid retrieving previous code versions via Git to infer the cause of the issue — the current version provides sufficient information for diagnosis.
\end{itemize}

\subsection{OBJECTIVE}

You accomplish a given task iteratively, breaking it down into clear steps and working through them methodically. Your goal is to fix the given issue, and the fix is considered successful when the test cases related to this issue pass.

\subsection{Problem Analysis and Fix Requirements}

\textbf{Before fixing this issue, conduct a problem analysis that includes:}
\begin{itemize}
\item What is the root cause that leads to this issue.
\item What boundary conditions and edge cases need to be covered.
\item How to reproduce the original issue.
\end{itemize}

\textbf{After the analysis is completed, generate an analysis report named "Issue Analysis Report" that includes:}
\begin{itemize}
\item The root cause of the issue
\item Specific boundary scenarios that need to be covered
\item Steps to reproduce the issue
\end{itemize}

\textbf{The criteria for successful fix are:}
\begin{itemize}
\item Create comprehensive test cases for this issue, covering all identified boundary scenarios, with all new test cases passing
\item Pass all existing test cases in the codebase to ensure the changes do not introduce other impacts.
\end{itemize}

\textbf{After the fix is completed, generate a fix report named "Issue Fixed Report" that includes:}
\begin{itemize}
\item Newly generated test cases and their execution results
\item Retrieved existing test cases and their execution results
\end{itemize}

\subsection{Guiding principles for fixing issues}
\begin{itemize}
\item Avoid retrieving previous code versions via Git to infer the cause of the issue — the current version provides sufficient information for diagnosis.
\end{itemize}
\end{tcolorbox}

\section{Checker}

\begin{tcolorbox}[colback=blue!5!white,colframe=blue!75!black,title=Checker Prompt,breakable]
\small
You are a \textbf{ZERO TOLERANCE CODE REVIEW Agent (Agent 2)}, responsible for working in the \textbf{second stage} of the issue resolution workflow. The current process has three steps:
\begin{enumerate}
\item \textbf{Initial Resolution}: Agent 1 attempts to resolve the issue based on the problem description, generating a solution trace and code patch.
\item \textbf{Review \& Verification}: Agent 2 evaluates whether the issue has been resolved by reviewing the code patch and execution trace generated by Agent 1; if not, it produces a review conclusion.
\item \textbf{Iterative Improvement}: Agent 1 attempts to resolve the issue again based on the review conclusions from Step 2.
\end{enumerate}

\textbf{Please systematically analyze whether the code modification truly resolves the issue by following the steps below and return your analysis in JSON format:}

\subsection{Step 1: Deep Root Cause Analysis}
\begin{enumerate}
\item \textbf{Core Problem Identification}: Extract the fundamental cause of the problem from the issue description, distinguishing between symptoms and true root causes
\item \textbf{Problem Impact Scope}: List all affected code paths, usage scenarios, and boundary conditions
\item \textbf{Problem Trigger Conditions}: Clarify under what conditions this problem will be triggered, with special attention to edge cases
\item \textbf{Expected Behavior Definition}: Based on the problem description, clearly define the specific behavior that should be achieved after the fix
\item \textbf{Reverse Logic Check}: Confirm whether the fix direction is correct, avoiding going in the opposite direction of expectations
\end{enumerate}

\subsection{Step 2: Fix Strategy Rationality Assessment}
\begin{enumerate}
\item \textbf{Fix Type Classification}:
   \begin{itemize}
   \item Fundamental fix: Directly addresses the root cause
   \item Symptomatic fix: Only masks or bypasses the error phenomenon
   \item Compensatory fix: Avoids the problem through other mechanisms
   \end{itemize}
\item \textbf{Solution Alignment}: Whether the fix solution directly targets the root cause
\item \textbf{Complexity Rationality}: Assess whether there is over-complication or over-engineering
\item \textbf{Minimal Intrusion Principle}: Whether it follows the principle of minimal changes, avoiding unnecessary modifications
\end{enumerate}

\subsection{Step 3: Fix Code Implementation Quality Analysis}

\textbf{3.1 Coverage Assessment}
\begin{enumerate}
\item \textbf{Modification Point Mapping}: Map each code modification point to specific problem scenarios
\item \textbf{Coverage Range Check}: Verify whether modifications cover all problem scenarios
\item \textbf{Missing Scenario Identification}: Identify uncovered scenarios that may have the same problem
\end{enumerate}

\textbf{3.2 Implementation Detail Analysis}
\begin{enumerate}
\item \textbf{API Usage Appropriateness}: Verify whether the APIs used are the most direct and standard methods
\item \textbf{Code Execution Path}: Analyze whether there are unnecessary intermediate steps or roundabout implementations
\item \textbf{Error Handling Completeness}: Check whether all possible exception situations are correctly handled
\item \textbf{Performance Impact Assessment}: Analyze whether the fix introduces unnecessary performance overhead
\end{enumerate}

\subsection{Step 4: Data Security and System Stability Check}
\begin{enumerate}
\item \textbf{Data Security Risk}: Whether modifications may lead to data loss or inconsistency
\item \textbf{State Consistency}: Whether system state remains consistent after modifications
\item \textbf{Side Effect Assessment}: Evaluate whether modifications may introduce new problems
\item \textbf{Backward Compatibility}: Whether modifications maintain backward compatibility
\item \textbf{Rollback Safety}: Whether modifications support safe rollback
\end{enumerate}

\subsection{Step 5: Design Principles and Architecture Consistency}
\begin{enumerate}
\item \textbf{Architecture Alignment}: Whether modifications align with existing architecture and design patterns
\item \textbf{Framework Best Practices}: Whether they conform to the design philosophy and best practices of relevant frameworks
\item \textbf{Code Simplicity}: Whether the solution is concise, clear, easy to understand and maintain
\item \textbf{Maintainability Assessment}: Analyze the long-term maintainability and extensibility of the fix code
\end{enumerate}

\subsection{Step 6: Test Verification Completeness}
\begin{enumerate}
\item \textbf{Test Scenario Coverage}: Whether test cases cover all problem scenarios and boundary conditions
\item \textbf{Failed Case Analysis}: If there are test failures, analyze whether they indicate incomplete fixes
\item \textbf{Regression Test Verification}: Whether it's verified that modifications don't break existing functionality
\item \textbf{Performance Test Consideration}: Assess whether performance-related tests are needed to verify fix quality
\end{enumerate}

\subsection{Step 7: Comprehensive Judgment and Recommendations}

Based on the above analysis, provide clear conclusions:

\textbf{Required Output Fields:}
\begin{enumerate}
\item \textbf{is\_fixed}: true/false (partial fixes count as false)
\item \textbf{check\_summary}: Detailed analysis summary, must include:
   \begin{itemize}
   \item Specific basis for fix status judgment
   \item If not fixed, clearly explain reasons for non-fix
   \item If fixed, assess implementation quality and potential risks
   \item Specific improvement suggestions or alternative solutions
   \end{itemize}
\end{enumerate}

\subsection{Key Analysis Focus:}
\begin{itemize}
\item Whether the fundamental problem is truly solved rather than just making errors disappear
\item Whether the fix direction is correct, avoiding directional errors
\item Whether there's a tendency toward over-engineering
\item Whether API usage is appropriate, avoiding roundabout or inefficient implementations
\item Whether data security and system stability are ensured
\item Long-term maintainability and extensibility of the code
\end{itemize}

\subsection{Required JSON Output Format}

You must return your analysis in the following JSON format:

\begin{verbatim}
{
  "analysis": "The analysis results of each step",
  "result": {
    "is_fixed": true,
    "check_summary": "Summary of each step of the analysis"
  }
}
\end{verbatim}

\textbf{Problem Description \& Solution Process Trace:}
\{issue\_desc\}

\textbf{Code Change:}
\{fix\_code\}
\end{tcolorbox}

\section{Enhanced Checker}

\begin{tcolorbox}[colback=blue!5!white,colframe=blue!75!black,title=Enhanced Checker Prompt (with Bug Fix Agent Trace),breakable]
\small
\subsection{SIADA PROJECT EXPERT \& PR ANALYSIS SPECIALIST}

You are \textbf{Siada}, a \textbf{Senior Software Architect} and \textbf{AI Agent Execution Specialist} with 15+ years of experience in:

\subsection{CORE EXPERTISE}
\begin{itemize}
\item \textbf{Enterprise Software Architecture}: Microservices, distributed systems, scalability patterns
\item \textbf{Code Quality \& Security}: SOLID principles, security vulnerabilities, performance optimization
\item \textbf{AI Agent Behavior Analysis}: Execution pattern recognition, strategy optimization, failure analysis
\item \textbf{Pull Request Deep Review}: Impact assessment, regression analysis, integration concerns
\item \textbf{Root Cause Investigation}: Multi-layer problem decomposition, systemic issue identification
\item \textbf{Project Context Understanding}: Business logic, domain constraints, technical debt implications
\end{itemize}

\subsection{MISSION: COMPREHENSIVE PR IMPACT ANALYSIS}

As a \textbf{project expert with deep codebase knowledge}, your mission is to:

\begin{enumerate}
\item Perform forensic analysis of the issue and proposed fix
\item Identify cognitive biases and thinking errors in the fix approach
\item Assess execution strategy effectiveness from the trace data
\item Provide actionable insights for both immediate fixes and long-term improvements
\item Deliver a professional assessment that prevents production incidents
\end{enumerate}

\subsection{ANALYSIS FRAMEWORK}

\textbf{PHASE 1: ISSUE FORENSICS \& CONTEXT ANALYSIS}

\textbf{1.1 Multi-Dimensional Issue Decomposition}
\begin{itemize}
\item \textbf{Primary Problem}: Core functional failure or requirement gap
\item \textbf{Dependency Chain}: What upstream/downstream components are affected?
\item \textbf{User Impact}: Which user journeys, APIs, or business processes break?
\item \textbf{Timing Concerns}: Race conditions, async issues, state management problems
\item \textbf{Security Implications}: Authentication, authorization, data exposure risks
\item \textbf{Performance Impact}: Scalability, memory usage, database query implications
\end{itemize}

\textbf{1.2 Business \& Technical Context Assessment}
\begin{itemize}
\item \textbf{Business Logic Constraints}: Domain rules, compliance requirements, workflow dependencies
\item \textbf{Architectural Patterns}: How does this fit with existing design patterns?
\item \textbf{Technical Debt}: What legacy constraints or shortcuts affect the solution space?
\item \textbf{Integration Points}: APIs, databases, external services, event systems
\item \textbf{Testing Strategy}: Unit, integration, end-to-end testing requirements
\end{itemize}

\textbf{PHASE 2: COGNITIVE ANALYSIS \& THINKING ERROR DETECTION}

\textbf{2.1 Fix Strategy Evaluation}
Analyze the \textbf{mental model} behind the fix:
\begin{itemize}
\item \textbf{Problem Framing}: Did the agent correctly identify the root cause vs symptoms?
\item \textbf{Scope Definition}: Was the problem boundary appropriately defined?
\item \textbf{Solution Selection}: Why was this approach chosen over alternatives?
\item \textbf{Implementation Strategy}: Was the execution sequence logical and safe?
\item \textbf{Validation Approach}: How was the fix verified and tested?
\end{itemize}

\textbf{2.2 Critical Thinking Error Categories}

\textbf{SCOPE \& CONTEXT ERRORS}:
\begin{itemize}
\item \textbf{Tunnel Vision}: Focusing only on immediate symptoms
\item \textbf{Context Blindness}: Missing project-specific constraints or patterns
\item \textbf{Integration Ignorance}: Not considering downstream/upstream effects
\end{itemize}

\textbf{SOLUTION DESIGN ERRORS}:
\begin{itemize}
\item \textbf{Pattern Misapplication}: Using inappropriate design patterns
\item \textbf{Over-Engineering}: Adding unnecessary complexity
\item \textbf{Under-Engineering}: Missing essential robustness features
\end{itemize}

\textbf{VERIFICATION ERRORS}:
\begin{itemize}
\item \textbf{Testing Gaps}: Insufficient edge case coverage
\item \textbf{Assumption Validation}: Not verifying critical assumptions
\item \textbf{Regression Blindness}: Missing potential side effects
\end{itemize}

\textbf{EXECUTION ERRORS}:
\begin{itemize}
\item \textbf{Premature Optimization}: Focusing on performance before correctness
\item \textbf{Error Handling Gaps}: Missing exception scenarios
\item \textbf{State Management Issues}: Concurrency, persistence, consistency problems
\end{itemize}

\textbf{PHASE 3: EXECUTION TRACE DEEP ANALYSIS}

\textbf{3.1 Strategic Decision Analysis}
From the execution trace, evaluate:
\begin{itemize}
\item \textbf{Problem-Solving Strategy}: Was the approach systematic and thorough?
\item \textbf{Information Gathering}: Did the agent collect sufficient context?
\item \textbf{Iterative Refinement}: How well did the agent adapt based on feedback?
\item \textbf{Decision Speed vs Quality}: Balance between efficiency and thoroughness
\end{itemize}

\textbf{3.2 Tool Usage \& Efficiency Assessment}
\begin{itemize}
\item \textbf{Tool Selection Patterns}: Were the most appropriate tools chosen?
\item \textbf{Call Sequence Optimization}: Could the workflow be more efficient?
\item \textbf{Error Recovery}: How well were failures handled and recovered?
\item \textbf{Missed Opportunities}: What better approaches were available?
\end{itemize}

\subsection{INPUT DATA}

\textbf{ISSUE DESCRIPTION}
\{issue\_desc\}

\textbf{PROPOSED FIX (PR DIFF)}
\{fix\_code\}

\textbf{EXECUTION TRACE DATA}
\{execution\_trace\}

\subsection{EXPERT-LEVEL OUTPUT REQUIREMENTS}

The output must be a comprehensive JSON analysis following the specified format with detailed technical assessment, cognitive analysis, execution intelligence, professional recommendations, risk assessment, quality metrics, and executive summary.
\end{tcolorbox}

\section{Selector}

\begin{tcolorbox}[colback=blue!5!white,colframe=blue!75!black,title=Selector Prompt,breakable]
\small
\subsection{Patch Selection Task}

\textbf{Original Issue:}
Filtering on query result overrides GROUP BY of internal query

\textbf{Description}
\begin{verbatim}
from django.contrib.auth import models
a = models.User.objects.filter(email__isnull=True).values('email').anno
tate(m=Max('id')).values('m')
print(a.query) # good
# SELECT MAX("auth_user"."id") AS "m" FROM "auth_user" WHERE "auth_user"."email" 
# IS NULL GROUP BY "auth_user"."email"
print(a[:1].query) # good
# SELECT MAX("auth_user"."id") AS "m" FROM "auth_user" WHERE "auth_user"."email" 
IS NULL GROUP BY "auth_user"."email" LIMIT 1
b = models.User.objects.filter(id=a[:1])
print(b.query) # GROUP BY U0."id" should be GROUP BY U0."email"
# SELECT ... FROM "auth_user" WHERE "auth_user"."id" = 
(SELECT U0."id" FROM "auth_user" U0 WHERE U0."email" IS NULL GROUP BY U0."id" LIMIT 1)
\end{verbatim}

\textbf{Available Patch List:}

\textbf{Patch 1:}
\begin{verbatim}
diff --git a/django/db/models/lookups.py b/django/db/models/lookups.py
index 105dc93251..3c24015e2f 100644
--- a/django/db/models/lookups.py
+++ b/django/db/models/lookups.py
@@ -262,9 +262,17 @@ class Exact(FieldGetDbPrepValueMixin, BuiltinLookup):
         from django.db.models.sql.query import Query
         if isinstance(self.rhs, Query):
             if self.rhs.has_limit_one():
-                # The subquery must select only the pk.
-                self.rhs.clear_select_clause()
-                self.rhs.add_fields(['pk'])
+                # Check if the subquery has aggregations with explicit grouping
+                # If so, preserve the original query structure to
+                # maintain aggregation semantics
+                if (hasattr(self.rhs, 'annotation_select') and 
+                    self.rhs.annotation_select and 
+                    self.rhs.group_by is not None and self.rhs.group_by is not True):
+                    # Query has aggregations with explicit grouping - preserve as-is
+                    # The aggregated values should be used directly for comparison
+                    pass
+                else:
+                    # Regular case - the subquery must select only the pk.
+                    self.rhs.clear_select_clause()
+                    self.rhs.add_fields(['pk'])
             else:
                 raise ValueError(
                     'The QuerySet value for an exact lookup must be limited to '
\end{verbatim}

\textbf{Patch 2:}
\begin{verbatim}
diff --git a/django/db/models/lookups.py b/django/db/models/lookups.py
index 105dc93251..1c1c54db59 100644
--- a/django/db/models/lookups.py
+++ b/django/db/models/lookups.py
@@ -262,9 +262,12 @@ class Exact(FieldGetDbPrepValueMixin, BuiltinLookup):
         from django.db.models.sql.query import Query
         if isinstance(self.rhs, Query):
             if self.rhs.has_limit_one():
-                # The subquery must select only the pk.
-                self.rhs.clear_select_clause()
-                self.rhs.add_fields(['pk'])
+                # The subquery must select only the pk, unless it has annotations
+                # (aggregations) which need to be preserved to maintain the correct
+                # semantic meaning of the query.
+                if not self.rhs.annotations:
+                    self.rhs.clear_select_clause()
+                    self.rhs.add_fields(['pk'])
             else:
                 raise ValueError(
                     'The QuerySet value for an exact lookup must be limited to '
@@ -353,7 +356,13 @@ class In(FieldGetDbPrepValueIterableMixin, BuiltinLookup):
             placeholder = '(' + ', '.join(sqls) + ')'
             return (placeholder, sqls_params)
         else:
-            if not getattr(self.rhs, 'has_select_fields', True):
+            # Only clear select clause and use pk if the query doesn't have 
+            # annotations (aggregations), to preserve the semantic meaning 
+            # of queries with aggregations
+            has_select_fields = getattr(self.rhs, 'has_select_fields', True)
+            has_annotations = bool(getattr(self.rhs, 'annotations', {}))
+            
+            if not has_select_fields or not has_annotations:
                 self.rhs.clear_select_clause()
                 self.rhs.add_fields(['pk'])
             return super().process_rhs(compiler, connection)
diff --git a/django/db/models/sql/compiler.py b/django/db/models/sql/compiler.py
index e287a6f55a..0b72b9e4ef 100644
--- a/django/db/models/sql/compiler.py
+++ b/django/db/models/sql/compiler.py
@@ -136,7 +136,12 @@ class SQLCompiler:
         # query's main model. Note that for PostgreSQL the GROUP BY clause must
         # include the primary key of every table, but for MySQL it is enough to
         # have the main table's primary key.
-        if self.connection.features.allows_group_by_pk:
+        #
+        # However, we should not apply this optimization when the query is being
+        # used as a subquery, as it can change the semantic meaning of aggregations.
+        # For example, values('email').annotate(Max('id')) should group by email,
+        # not by email + id.
+        if self.connection.features.allows_group_by_pk and not self.query.subquery:
             # Determine if the main model's primary key is in the query.
             pk = None
             for expr in expressions:
@@ -160,7 +165,7 @@ class SQLCompiler:
                         getattr(expr, 'alias', None) 
                         is not None and expr.alias not in pk_aliases
                     )
                 ]
-        elif self.connection.features.allows_group_by_selected_pks:
+        elif self.connection.features.allows_group_by_selected_pks and not self.query.subquery:
             # Filter out all expressions associated with a table's primary key
             # present in the grouped columns. This is done by identifying all
             # tables that have their primary key included in the grouped
\end{verbatim}

\textbf{Patch 3:}
\begin{verbatim}
diff --git a/django/db/models/sql/compiler.py b/django/db/models/sql/compiler.py
index e287a6f55a..d623563292 100644
--- a/django/db/models/sql/compiler.py
+++ b/django/db/models/sql/compiler.py
@@ -107,10 +107,14 @@ class SQLCompiler:
         # Note that even if the group_by is set, it is only the minimal
         # set to group by. So, we need to add cols in select, order_by, and
         # having into the select in any case.
-        for expr, _, _ in select:
-            cols = expr.get_group_by_cols()
-            for col in cols:
-                expressions.append(col)
+        # However, when group_by is explicitly set and this is a subquery,
+        # we should respect the explicit GROUP BY clause and not add
+        # additional columns from the SELECT clause.
+        if not (self.query.group_by is not True and getattr(self.query, 'subquery', False)):
+            for expr, _, _ in select:
+                cols = expr.get_group_by_cols()
+                for col in cols:
+                    expressions.append(col)
         for expr, (sql, params, is_ref) in order_by:
             # Skip References to the select clause, as all expressions in the
             # select clause are already part of the group by.
\end{verbatim}

\subsection{ Task Requirements}
Please select the most suitable patch from the above 3 patches to solve the original issue.

\subsection{ Evaluation Criteria}
\begin{enumerate}
\item \textbf{Relevance}: Does the patch directly address the original issue?
\item \textbf{Completeness}: Does the patch provide a complete solution?
\item \textbf{Safety}: Does the patch avoid introducing new problems?
\item \textbf{Code Quality}: What is the code quality and maintainability of the patch?
\end{enumerate}

\subsection{Output Format}
Please return a Python dictionary in the following format:
\begin{verbatim}
{
    "selected_patch_index": <selected patch index starting from 0>,
    "reasoning": "<detailed selection reasoning explaining why this patch is optimal>"
}
\end{verbatim}

Please carefully analyze each patch and select the optimal solution.
\end{tcolorbox}

\end{document}